\documentstyle[preprint,12pt,aps,eqsecnum,tighten]{revtex}

\newcommand{\beq}{\begin{equation}}
\newcommand{\eeq}{\end{equation}}
\newcommand{\beqa}{\begin{eqnarray}}
\newcommand{\eeqa}{\end{eqnarray}}
\newcommand{\Adir}{\mbox{${\cal A}^{\rm dir}_{\rm CP}$}}
\newcommand{\bBz}{\mbox{${\overline{B^0}}$}}
\newcommand{\bKz}{\mbox{${\overline{K^0}}$}}

\begin{document}

\draft

{\tighten

\preprint{
\vbox{\hbox{JHU--TIPAC--97018}
      \hbox{UCHEP-79}
      \hbox{WIS-97/36/Dec-PH}
      \hbox{hep-ph/9712225}
      \hbox{December 1997} }}

\title{Final State Interactions and New Physics in $B\to\pi K$ Decays}
\author{Adam F. Falk${}^{a}$, Alexander L. Kagan${}^{b}$,
Yosef Nir${}^{c}$ and Alexey A. Petrov${}^{a}$}
\address{ \vbox{\vskip 0.truecm}
${}^a$ Department of Physics and Astronomy, The Johns Hopkins University, \\
       3400 North Charles Street, Baltimore, Maryland 21218 USA \\
${}^b$ Department of Physics, \\ University of Cincinnati,
       Cincinnati, Ohio 45221 USA \\
${}^c$ Department of Particle Physics, \\
       Weizmann Institute of Science, Rehovot 76100, Israel}

\maketitle
\thispagestyle{empty}
\setcounter{page}{0}
\begin{abstract}%
Within the Standard Model, and if one assumes that soft rescattering
effects are negligible, the CP asymmetry $\Adir(B^\pm\to\pi^\pm K)$
is predicted to be very small and the ratio
$R = BR(B_d \to \pi^\mp K^\pm)/BR(B^\pm \to \pi^\pm K)$
provides a bound on the angle $\gamma$ of the unitarity triangle,
$\sin^2\gamma\leq R$. We estimate the corrections from soft rescattering
effects using an approach based on Regge phenomenology, and find effects of
order $10\%$ with large uncertainties. In particular, we conclude that
$\Adir\sim0.2$ and $\sin^2\gamma\sim1.2R$ could not be taken
unambiguously to signal New Physics. Using $SU(3)$ relations, we suggest
experimental tests that could constrain the size of the soft rescattering effects
thus reducing the related uncertainty. Finally, we study the effect of various
models of New Physics on $\Adir$ and on $R$.

\end{abstract}

} % end tighten

\newpage

%%%%%%%%%%%%%%%%%%%%%%%%%%%%%%%%%%%%%
\section{Introduction}
Heavy quark decays serve as a powerful tool for
testing the Standard Model and provide invaluable
possibilities to study CP violation.  However, the
interpretation of experimental observables in terms of fundamental parameters
is often less than clear. Rare hadronic decays of $B$ mesons, for
example, proceed through both tree level Cabibbo-suppressed amplitudes and
through one loop penguin amplitudes. On the one hand, this situation allows direct
CP violating effects that may give the first evidence for CP violation outside the
neutral kaon system. On the other, these competing contributions complicate
the extraction of Cabibbo-Kobayashi-Maskawa (CKM) angles and, in particular,
the angle $\gamma$,
\beq\label{defgamma}
\gamma\equiv\arg\left[-{V_{ud}V_{ub}^*\over V_{cd}V_{cb}^*}\right].
\eeq

The CLEO collaboration has presented combined
branching ratios for $B^\pm\to\pi^\pm K$ and $B_d \to \pi^\mp K^\pm $
\cite{CLEOBR} making these modes of particular interest.  In the Standard Model,
these decays are mediated by the $\Delta B = 1$ Hamiltonian, which takes the form
\beqa\label{HeffCP} {\cal H}_{\rm eff} &=& \frac{G_F}{\sqrt{2}}\left[
 V_{cb} V_{cs}^* \left(\sum_{i=1}^{2}
C_i Q_i^{c s} + \sum_{i = 3}^6 C_i Q_i^s + \sum_{i = 7}^{10}
C_i Q^s_i \right)\right.\nonumber\\
&&\qquad\left.\mbox{}+ V_{ub} V_{us}^* \left(\sum_{i=1}^{2}
C_i Q_i^{u s} + \sum_{i = 3}^6 C_i Q_i^s + \sum_{i = 7}^{10}
C_i Q^s_i \right)\right] + {\rm H.c.}\,. \eeqa
The flavor structures of the current-current, QCD penguin, and electroweak
penguin operators are, respectively,
$Q_{1,2}^{qs} \sim \bar s q  \bar q b $,
$Q_{3,..,6}^{s} \sim \bar s b \sum \bar q' q' $, and
$Q_{7,..,10}^{s} \sim \bar s b \sum e_{q'} \bar q' q' $, where the sum is over
light quark flavors~\cite{burasQi}.  The Wilson coefficients $C_i$ are
renormalization scale dependent; at a low scale $\mu \sim 1\,$GeV, they roughly
satisfy
$C_{1,2} = {\cal O}(1)$, $C_{3,..,6,9} = {\cal O}(10^{-2})$, and
$C_{7,8,10} \le {\cal O}(10^{-3} )$.  In many extensions of the Standard Model,
the effect of New Physics at low energies is simply to modify the values of the
Wilson coefficients.

A {\it scale-independent\/} way of decomposing the decay
amplitudes of interest is to do so not naively according to contributions of the
operators
$Q_i^{qs}$, but rather according to their dependence on the elements of the CKM
matrix,
\beqa \label{Camps}
A(B^+ \to \pi^+ K^0) = A_{cs}^+  - A_{us}^+ e^{i \gamma} e^{i \delta_+},~~~~~
A(B^- \to \pi^- \bKz) = A_{cs}^+ - A_{us}^+ e^{-i \gamma} e^{i \delta_+},
\nonumber\\
A(B^0 \to \pi^- K^+) =  A_{cs}^0 - A_{us}^0 e^{i \gamma} e^{i
\delta_0},~~~~~ A(\bBz \to \pi^+ K^-) =   A_{cs}^0
- A_{us}^0 e^{-i \gamma} e^{i \delta_0},
\eeqa
where $\delta_0$ and $\delta_+ $ are CP-conserving phases induced by the strong
interaction, and the dependence on the CKM phases is shown
explicitly. The first and second terms in each amplitude correspond to matrix
elements of the  first and second terms in ${\cal H}_{eff}$ (or their Hermitian
conjugates), respectively. Note that each term is by itself scheme and
renormalization scale  independent.

We will avoid, as much as possible, the common terminology of ``tree'' versus
``penguin'' contributions, which can lead to much unnecessary confusion.  The
standard convention is to take ``tree'' contributions to a given decay to be those
which are mediated by the current-current operators $Q_{1,2}^{qs}$, and ``QCD
penguin'' contributions to be those mediated by $Q_{3,\dots,6}^{qs}$.  However, this
is not a scale-invariant decomposition.  If the computation of physical matrix
elements could be accomplished perturbatively, this would not be such a serious
failing, as the scale-dependence of the matrix elements would cancel explicitly
against that of the Wilson coefficients.  But this is not the case; rather, the
matrix elements must be modeled phenomenologically, and the manifest
scale-independence of the result is lost.  The greatest difficulty is found when one
is interested, as we will be, in a significant contribution to some process which
arises essentially at long distances, where the the physics is intrinsically
nonperturbative.  There one can dispute endlessly, and pointlessly, about whether
what one is computing is ``really'' tree or penguin in nature.  The question
obviously has no unique answer, but its resolution is, fortunately, of no
practical consequence.

That having been said, one can still make some general statements about the
expected relative contributions of the operators in ${\cal H}_{\rm eff}$ to a given
exclusive decay mode.  The electroweak penguin operators are commonly neglected,
since the contributions with a sizable Wilson coefficient, $C_9Q^s_9$, are color
suppressed or require rescattering from intermediate states.  In this case isospin
symmetry of the strong interactions leads to the  simplification
$A_{cs}^0 =  A_{cs}^+ $.
It is now believed that the current-current operator
contributions to $A_{cs}^{0,+}$ are roughly of same order as the
QCD penguin operator contributions \cite{bf,ciuchini}.
To be specific, this observation is based on the fact that at next-to-leading order
in QCD perturbation theory, it holds for the corresponding parton model decays
$b \to u\bar u s$ and $b\to d \bar d s $.
The contribution of the current-current operators to
$A_{us}^0$ is also expected to be of the same order, despite the CKM
suppression,
because of the large value of $C_2 $, namely, $ V_{ub} V^*_{us}\, C_2
\sim V_{cb} V_{cs}^*\, C_{3,..6} $.
However, since for $B^\pm \to \pi^\pm K $ the relevant quark transition is
$b \to d \bar d s$,  one might expect the size of
$A_{us}^+ $ relative
to $A_{cs}^+ $ to be highly suppressed by the
small ratio $|V_{ub}V_{us}^* /V_{cb}V_{cs}^* |\sim 0.02$.  This would hold equally
for the current-current and penguin operators. If, indeed, $r_+ = A_{us}^+
/A_{cs}^+ \sim |V_{ub}V_{us}^* / V_{cb}V_{cs}^* |$  is a good approximation, then
there are two important consequences:

$(i)$ Direct CP violation could be observed, in principle, through
the CP asymmetry $\Adir\equiv\Adir(B^+\to\pi^+K^0)$,
\beq\label{ABpiK}
\Adir={
BR(B^+\to\pi^+K^0)-BR(B^-\to\pi^-\bKz)\over
BR(B^+\to\pi^+K^0)+BR(B^-\to\pi^-\bKz)}
={2r_+\sin\gamma\sin\delta_+\over1-2r_+\cos\gamma\cos\delta_++r_+^2}\,.
\eeq
However, it would be small,
\beq\label{smallACP}
\Adir(B^+\to\pi^+ K^0)\leq{\cal O}(\lambda^2),
\eeq
where $\lambda \simeq0.22 $ is the Wolfenstein parameter.
For example, ``hard'' final-state interaction estimates~\cite{GeHo,wyler,Flea,KPS},
where the $u$ quarks in $Q_{1,2}^{us} $ are treated as a
perturbative loop,
give
$\Adir \sim 1\% $.

$(ii)$ Model-independent bounds could be
obtained for the angle $\gamma$ using only the {\it combined\/}
branching ratios  $BR(B^\pm \to \pi^\pm K)$ and
$BR(B_d \to \pi^\mp K^\pm)$ \cite{fm,GNPS}.
One can construct the ratio
\begin{equation} \label{ratio}
R = \frac{BR(B^0 \to \pi^- K^+) + BR(\bBz \to \pi^+ K^-)}{
BR(B^+ \to \pi^+ K^0) + BR(B^- \to \pi^- \overline{K^0})}=
\left({A_{cs}^0\over A_{cs}^+}\right)^2{1-2r_0\cos\gamma\cos\delta_0+r_0^2
\over1-2r_+\cos\gamma\cos\delta_++r_+^2}\,,
\end{equation}
where $r_0=A_{us}^0 /A_{cs}^0$.
Assuming that $A_{us}^+ $ and electroweak
penguin operator contributions are negligible,
the ratio~(\ref{ratio}) takes the simple form
\begin{equation} \label{theory}
R = 1 - 2 r_0 \cos \gamma \cos \delta_0 + r_0^2\,.
\end{equation}
The observable $R$ may be minimized with respect to the
unknown hadronic parameter $r_0$, yielding the inequality
\begin{equation}
R \geq 1-\cos^2 \gamma \cos^2 \delta_0\,.
\end{equation}
Since $\cos^2\delta_0\le1$, this leads to the bound
\begin{equation} \label{bound}
\sin^2 \gamma  \leq R.
\end{equation}
The bound including electroweak penguin operators is obtained by
substituting
\beq\label{EWsub} R \to R (A_{cs}^+ /A_{cs}^0 )^2 \,.\eeq
It has recently been observed that the modified
bound could differ by as much as $\pm 10\% $ \cite{rosnergronau}.
If true, a stringent bound on $\gamma$ would be obtained if the
experimental errors in the presently reported $R_{\rm exp} = 0.65 \pm
0.40$~\cite{CLEOBR} were to be reduced, with the central value unchanged.

The $B\to\pi K$ transitions are suppressed in the Standard Model by
either CKM matrix elements or small Wilson coefficients. As a consequence, these
decays are potentially sensitive to New Physics. In particular, in the
presence of New Physics, a large CP asymmetry can be induced,
thus violating the bound (\ref{smallACP}),
and $R$ can be modified in a way that violates the bound (\ref{bound}).
The analyses leading to (\ref{smallACP}) and to (\ref{bound}),
however, explicitly assume that the CKM angle $\gamma$
does not enter the theoretical expression for the charged
decay amplitudes (\ref{Camps}).  As it is usually expressed, one requires the
absence of significant contributions from the current-current operators
$Q_{1,2}^{us}$ to this decay channel.  As noted above, this assumption is
based on the observation that the quark level decay $b\to d\bar ds$ is not mediated
direcly by $Q_{1,2}^{us}$.  However, this naive treatment of the dynamics ignores
the effects of soft rescattering effects at long distances, which can include the
exchange of global quantum numbers such as charge and strangeness.
It is the purpose of the next section to discuss an explicit model for such a
rescattering, and to consider its effect on the bounds~(\ref{smallACP})
and~(\ref{bound}).  Before presenting the model, however, we must make some general
comments about what we do, and what we do {\it not,} expect to accomplish with this
exercise.

We will consider the contributions to $B^+\to\pi^+K^0$ from rescattering through
coupled channels such as $B^+\to\pi^0(\eta)K^+$ or corresponding multi-body decays.
In fact, we will treat only the two body intermediate states explicitly.  The model
we will employ will be based on Regge phenomenology, including the exchange of the
$\rho$ trajectory and others related by $SU(3)$ flavor symmetry.  The coupling of the
trajectory to the final state will be extracted from data on $\pi p$ and $pp$
scattering cross sections.  A few points are in order.  First, we have little
confidence in the quantitative predictions of the model {\it per se.}  In fact, we
believe that it would be irresponsible to claim an accuracy of better than a factor
of two for the size of soft rescattering effects, using {\it any\/} model currently
available.  Neither our nor any other model should be taken as a canonical
framework for the estimate of final state interactions in $B$ decays.  Rather, the
purpose of our calculation is to be illustrative: our model will predict
$r_+$ at the level of ten percent, with no fine tuning or unnatural enhancements.
We will use this result to argue that such a value of $r_+$ is entirely generic
within the Standard Model.  At such a level, the effect of final state interactions
on total branching fractions is small, but the effect on quantities which are most
interesting for $r_+=0$, such as $\Adir(B^\pm\to\pi^\pm K)$ and $R$, can be more
dramatic.

Second, the rescattering effects which we will consider are {\it not\/} already
included in an analysis of the strong ``BSS'' phases induced when the virtual
particles in a penguin loop go onto the mass shell~\cite{bss}.  The fact that such
an analysis typically yields small corrections cannot be used to argue that
rescattering contributions to $r_+$ must be small.  On the contrary, we will
consider
intermediate states with on-shell pseudoscalar mesons, rather than on-shell
quarks.  In the absence of an argument that parton-hadron duality should hold in
exclusive processes involving pions and kaons (for which there is scant evidence),
one must conclude that the long distance physics of meson rescattering is not
probed by the BSS analysis.  (The recent proposal that rescattering effects must be
small~\cite{BFM} does {\it not\/} go outside the BSS framework in obtaining
quantitative estimates.)

Third, the issue of whether the processes we consider are of the ``tree'' or
``penguin'' type is a dangerous red herring.  As discussed above, the question has
no scale-invariant meaning, and also no important implications.  Our model
addresses contributions to the well-defined amplitude $A^+_{us}$.  Within the
model, we will first use the current-current operators $Q_{1,2}^{us}$ to generate
the transition $B\to\pi^0(\eta)K^-$, which will then rescatter to $\pi^-\overline{K^0}$. However, we will deliberately refrain from referring to this
as a ``tree'' contribution, in order to avoid unnecessary confusion.

Fourth, if one decomposes the matrix element for $B^-\to\pi^- \overline{K^0}$ into
amplitudes of definite isospin, the rescattering process which we will consider
is non-trivially embedded in their sum. Recent isospin analysis of this decay
~\cite{BFM,GW} have stressed the importance of the strong phases
associated with the isospin amplitudes. (For previous isospin analyses of
$B\to\pi K$ decays, see \cite{NiQu,LNQS}.) In the isospin language the magnitude of
$A_{us}^+$ depends on the differences between these phases and vanishes in
the limit that they vanish. However, the isospin decomposition sheds no light
on the sizes of these phase differences, hence on the magnitude of $A_{us}^+$,
leaving the former as free parameters. The literature presently contains a
variety of quite divergent opinions concerning the ``natural'' size of
rescattering effects in this decay~\cite{fm,rosnergronau,bss,BFM,GW}. The purpose of our calculation, however crude, is to address the issue more
quantitatively by providing the first analysis to model the {\it magnitude} of
rescattering in this decay. Finally, we note that the rescattering in question
is {\it inelastic,} despite its quasi-elastic kinematics, and cannot be studied
adequately in any model of purely {\it elastic\/} final state phases.

The calculation of the Standard Model predictions is given
in section 2. First, we describe how the final state interactions
(FSI) affect the CP asymmetry and the bound on $\gamma$.
Then we calculate, using the phenomenological Regge model,
FSI corrections for specific two body states.
In section 3 we suggest experimental tests that
could potentially give an upper bound on these contributions, independent of
hadronic models for the final state rescattering. In section 4
we analyze which types of New Physics models can significantly
affect the relevant $B\to\pi K$ decays, and whether there are
relations between such new contributions to the charged and
neutral modes. We summarize our results in section 5.

%%%%%%%%%%%%%%%%%%%%%%%%%%%%%%%%%%%%%%%%%%%%%%%%%%%%%
\section{Final State Interactions}

%%%%%%%%%%%%%%%%%%%%%%%%%%%%%%%%%%%%%%%%%%%%%%%%%%%%%
\subsection{The effects of FSI Corrections}

We would like to investigate the impact of final state rescattering on the CP
asymmetry $\Adir(B^\pm\to\pi^\pm K)$ and the ratio of branching fractions $R$.  The
rescattering   process involves an intermediate on-shell state $X$, such that $B\to
X\to K\pi$.  In particular, we assume that there exists a generic (multibody) state
$K n\pi$. In a straightforward generalization of Eq.~(\ref{Camps}),
the charged and neutral channel amplitudes can be written
as
\begin{eqnarray} \label{multibody}
A(B^+ \to K n\pi) &=& A_{cs}^{n+}  - A_{us}^{n+} e^{i \gamma}
e^{i \delta_+^n}\,,\\ \nonumber
A(B^0 \to K n\pi) &=& A_{cs}^{n0} - A_{us}^{n0} e^{i \gamma}
e^{i \delta_0^n}\,.
\end{eqnarray}
Rescattering contributions, again
decomposed according to their dependence on CKM factors,
are given by
\begin{eqnarray} \label{amplitudes}
A(B^+ \to K n\pi \to \pi^+ K^0) &=&
S_1^n A_{cs}^{n+} -
S_2^n A_{us}^{n+} e^{i \gamma}\,,\nonumber\\
A(B^0 \to K n\pi \to \pi^- K^+) &=&
S_3^n A_{cs}^{n0}
- S_4^n A_{us}^{n0} e^{i \gamma}\,,
\end{eqnarray}
where $S_i^n$ is the complex amplitude for rescattering from
a given multibody final state to the channel of interest.
Analogous contributions arise in the conjugated channels.
In the limit where one neglects electroweak penguin operator
contributions, isospin symmetry requires $A_{cs}^+=A_{cs}^0$, and this equality is
not spoiled by rescattering effects.  The $i=1,3,4$ rescattering amplitudes
can, for our purposes, be absorbed into the unknown amplitudes in
Eq.~(\ref{Camps}).

We are interested in the possibility that the rescattering of transitions mediated
by $Q_{1,2}^{us}$ are significant enough to dominate $A^+_{us}$, so we make the
approximation
\beq A_{us}^+ e^{i\delta_+} = \sum_n S_2^n A_{us}^{n+}\,,\eeq
and define $\epsilon=A_{us}^+/A_{cs}^+$.  Let us assume that rescattering effects
do not dominate the overall decay, so we may retain just terms linear in
$\epsilon$. In that case,
$\Adir$ of Eq.~(\ref{ABpiK}) and
$R$ of Eq.~(\ref{ratio}) take the form
\beq\label{corrACP}
\Adir
={2 \epsilon \sin\gamma\sin\delta_+
\over1-2 \epsilon \cos\gamma\cos\delta_+}\,,
\eeq
\beq \label{corrR}
R={1  - 2r_0 \cos \gamma \cos \delta_0 + r_0^2\over
1 - 2 \epsilon \cos \gamma \cos \delta_+ }\,.
\eeq
Once again, we may extremize $R$ with respect to the unknown $r_0$,
\begin{equation} \label{newbound1}
R \geq \frac{1 - \cos^2 \gamma \cos^2 \delta}
{1 - 2 \epsilon \cos \gamma \cos \delta_+}\,.
\end{equation}
Following the same line of reasoning as before with respect to the unknown strong
phases $\delta_0$ and $\delta_+$, we find the new bound
\beq\label{FMmod}
  \sin^2\gamma\le R(1+2\epsilon\sqrt{1-R}),
\eeq
or solving for $\cos \gamma$,
\begin{equation} \label{newbound2}
|\cos \gamma|\geq \sqrt{1-R}-\epsilon R\,.
\end{equation}

It is clear that
even a small rescattering amplitude $\epsilon \sim 0.1$ could induce
a significant shift in the bound on $\gamma$ deduced from $R$, effectively
diminishing the model-independent bound of Ref.~\cite{fm} as $R \to 1$.  It is also
clear from (\ref{corrACP}) that a small rescattering effect,  again
$\epsilon \sim 0.1$, could in principle
generate an ${\cal O}(10\%) $ CP asymmetry
which is significantly larger than the bound (\ref{smallACP}) on
$\Adir(B^\pm\to\pi^\pm K)$. Therefore, in order to understand whether a large CP
asymmetry signals New Physics, and whether it is possible to obtain a bound on
$\gamma$, it is useful to employ a particular model of soft FSI to obtain an
order of magnitude
estimate of the effect. In the next section we will estimate the amplitude
$\epsilon$ using a phenomenological approach based on the exchange of Regge
trajectories~\cite{fsi1,fsi2,fsi3}.

%%%%%%%%%%%%%%%%%%%%%%%%%%%%%%%%%%%%%%%%%%%%%
\subsection{The two body rescattering contribution}

We will estimate the contribution to $\epsilon$ from the rescattering
of certain two body intermediate states.  While these channels alone are not
expected to dominate rescattering~\cite{fsi1}, we might expect them to provide
a conservative lower bound on the size of the effect.  At any rate, it would
be peculiar for the total effect of rescatttering to be significantly {\it
smaller\/} than the two body contribution.  The most important channels in
charged $B$ decays that might rescatter to the final states of interest are
$B^- \to \pi^0 K^-$ and $B^- \to \eta K^-$.  (To be conservative, we will
neglect the $\eta' K$ channel, which is unrelated to the others by $SU(3)$
symmetry.  Its inclusion would likely enhance the effect which we will
find.)  To estimate the contribution to $\epsilon$ of these channels, it is
necessary to estimate both the relative amplitude $A_{us}^{2+}/A_{cs}^+$ for
producing the intermediate state, and the amplitude $\epsilon_2=|S_2|$ for it to
rescatter to the final state of interest.  In our model, then,
$\epsilon=\epsilon_2(A_{us}^{2+}/A_{cs}^+)$, and we will extract only the
magnitude $\epsilon$.  We will not attempt to predict the strong phase
$\delta_+$, which is even more model-dependent.

The amplitude ratio $A_{us}^{2+}/A_{cs}^+ $ may be estimated using
factorization~\cite{fact1,fact2,fact3,fact4} and the Bauer-Stech-Wirbel~\cite{bsw}
model, starting from  ${\cal H}_{eff} $.  We will ignore the electroweak penguin
operators. In addition to the soft FSI contributions which we will estimate
using the leading order Wilson coefficients, there are
also ``hard'' FSI phases which can be generated via quark rescattering~\cite{bss}
and are estimated at next-to-leading order~\cite{wyler}.  As discussed earlier,
the final state interactions which we consider are distinct from these ``BSS''
phases.

We will impose $SU(3)$ symmetry on all aspects of our estimate of $\epsilon$.
Corrections to the $SU(3)$ limit are typically at the level of 30\%, small
compared with other uncertainties in the calculation.  Hence we must consider
both the intermediate states $\pi^0 K^+$ and $\eta K^+$.  For normalization, we
will also need to compute $A_{cs}^+$, which in the BSW model is induced by the QCD
penguin operators.   The computation is straightforward, and we find
\begin{eqnarray} \label{ampsfact}
A^{2+}_{us}(B^+ \to P^0 K^+) = &-& G_F m_B^2 | V_{ub} V_{us}^*| \Biggl \{
\left( C_1 + C_2/3 \right) F_{P^0}^{uu}
f_+^K (m_{P^0}^2) L_K (\mu_{P^0})
\nonumber \\
&&\qquad\qquad+ \left( C_2 + C_1/3 \right) F_{K}^{su}
f_+^{P^0} (m_K^2) L_\pi (\mu_{P^0}) \Biggr \}
\nonumber
\\
A_{cs}^+ (B^+ \to \pi^+ K^0) = &-& G_F m_B^2 |V_{cb} V_{cs}^*| \Biggl \{
\left( C_4 + C_3/3 \right) F_{K^0}^{sd}
f_+^\pi (m_K^2) L_K (\mu_{\pi})
\nonumber \\
&&\qquad\qquad+ \left( C_6 + C_5/3 \right) F_{K^0}^{sd}
f_+^\pi (m_K^2) \frac{2 m_K^2}{m_s m_b} M_\pi (\mu_K)
\Biggr \},
\end{eqnarray}
for $P^0=\pi^0, \eta$.
The form factors $f^P_+(q^2)$, the decay constants
$F^{q_1q_2}_P$, and the kinematic functions $L_P(\mu_i)$ and $M_P(\mu_i)$  are
defined as follows:
\begin{eqnarray} \label{formfactors}
\langle P | \bar q \gamma^\mu b | B \rangle &=&
f_+^P (q^2) (p_B+p_P)^\mu + f_-^P (q^2) (p_B-p_P)^\mu\,,
\nonumber \\
\langle P | \bar q_1 \gamma^\mu \gamma_5 q_2 | 0 \rangle &=&
-i \sqrt{2}\, F_P^{q_1 q_2}\, p_P^\mu\,,\nonumber\\
L_P(\mu) &=& 1- \mu_P + \frac{f_-^P (q^2)}{f_+^P (q^2)}
\mu\,, \nonumber \\
M_P(\mu) &=& \frac{1}{2} \Biggl [(3-y+(1-3y)\mu_P -(1-y)\mu)
\nonumber \\
&&\qquad \mbox{}+ \frac{f_-^P (q^2)}{f_+^P (q^2)}
(1-y+(1-y)\mu_P +(1+y)\mu)
\Biggr],
\end{eqnarray}
where $\mu_i = m_i^2/m_B^2$, $q^2$ is the momentum carried away by the current, and
$y\approx 1/2-1$ is related to the distribution of quark momenta in the pseudoscalar
mesons.  Note that in the BSW model, the dominant contribution to
$A^{2+}_{us}(B^+ \to P^0 K^+)$ is from the current-current operators $Q^{us}_{1,2}$,
while the dominant contribution to $A^{2+}_{cs}(B^+\to\pi^+K^0)$ is from the QCD
penguins $Q^{cs}_{3,\dots,6}$.

The ratios $(A_{us}^{2+} /A^+_{cs} )_{\pi^0 , \eta} $ simplify
significantly if $SU(3)$ symmetry is imposed on the quantities which appear
in Eq.~(\ref{ampsfact}).  In particular, we make use of the $SU(3)$ relations
\begin{eqnarray} \label{su3rels}
F_{\pi^0}^{uu} = F_K^{sd}/\sqrt{2}\,,
&&\qquad f_+^{\pi^0} (0) =  f_+^K (0)/\sqrt{2}\,, \nonumber \\
F_{\eta}^{uu} = F_K^{sd}/\sqrt{6}\,,
&&\qquad f_+^{\eta} (0) =  f_+^K (0)/\sqrt{6}\,.
\end{eqnarray}
We have checked that the inclusion of $\eta-\eta'$ mixing, which violates
$SU(3)$, would  change our final answer by no more than 20\%.
The ratios of interest then may be written
\begin{eqnarray} \label{ratios}
\left({A_{us}^{2 +} \over A^+_{cs} }\right)_{\pi^0 }   &=&
\frac{|V_{ub} V_{us}^*|}{|V_{cb} V_{cs}^*|}\,
\frac{(1+1/3)(C_1 + C_2)/\sqrt2}
{\left (C_4 + C_3/3 \right) +
\left (C_6 + C_5/3 \right) (2 m_K^2/m_s m_b)
(M_\pi(\mu_K)/L_\pi(\mu_K))}\,, \nonumber \\
\left({A_{us}^{2 +} \over A^+_{cs} }\right)_{\eta } &=& \frac{1}{\sqrt{3}}\,
\left({A_{us}^{2 +} \over A^+_{cs} }\right)_{\pi^0 }\,,
\end{eqnarray}
where $\eta-\eta'$ mixing is neglected.  With values for the coefficients
$C_i$ taken from Ref.~\cite{BBL}, at leading order and with
$\Lambda^{(5)}_{\overline{\rm MS}}=225\,$MeV, we find
$|(A_{us}^{2+} /A^+_{cs} )_{\pi^0}|\simeq0.35$.

We now turn to an estimate of the rescattering amplitude $\epsilon_2$.  Our
technique is described in detail in Ref.~\cite{fsi1}, and here we only outline
the procedure.  We begin by writing an
expression for the discontinuity of the amplitude for the charged
$B$ decay,
\begin{eqnarray} \label{disc}
{\rm Disc} ~A (B^- \to \overline{K^0} \pi^-)_{\rm FSI} &=& \frac{1}{2}
\int \frac{d^3 {\bf p}_1}{(2 \pi)^3 2 E_1}
\frac{d^3 {\bf p}_2}{(2 \pi)^3 2 E_2}
(2 \pi)^4 \delta^{(4)} (p_B - p_1 - p_2)
\nonumber \\
&&\quad\times A (B^- \to \pi^0 (p_1) K^-(p_2))\,
{\cal M} (\pi^0 (p_1) K^-(p_2) \to \overline{K^0} \pi^-)\,.
\end{eqnarray}
The rescattering matrix element ${\cal M} (\pi^0 (p_1) K^-(p_2) \to\overline{K^0} \pi^-)$ has a well known parameterization inspired by Regge
phenomenology~\cite{fsi1}.  For the exchange of the leading $\rho$ trajectory,
it may be written as
\begin{eqnarray} \label{reggeme}
{\cal M}^+ (\pi^0K^- \to \overline{K^0} \pi^-) =
\gamma(t)\ \frac{e^{-i \pi\alpha (t)/2}}{
\cos (\pi\alpha (t)/2)}
\ \left ( \frac{s}{s_0} \right )^{\alpha (t)}\,,
\end{eqnarray}
where $\gamma(t)$ is a residue function,
$s$ and $t$ are the Mandelstam variables, and $s_0 = 1\,{\rm GeV}^2$ is an
arbitrary hadronic scale.  We take a linear Regge trajectory,
$\alpha(t)=\alpha_0+\alpha't$, with $\alpha_0=0.44$ and
$\alpha'=0.94\,{\rm GeV}^{-2}$.  Since $\alpha_0$ is approximately the same for
the $\rho$, $K^*$ and $\omega$ trajectories, it is convenient to
consider a single ``octet'' trajectory which carries a representation of $SU(3)$
and contains all the vector mesons.
Taking $\gamma(t)=\gamma(0)\equiv\gamma$ for simplicity, the discontinuity
can be calculated,
\begin{equation}
{\rm Disc} ~A (B^- \to \overline{K^0} \pi^-)_{\rm FSI} = \gamma\bar\epsilon_2(s)
\left( \frac{s}{s_0} \right )^{\alpha_0 - 1}
A (B^- \to \pi^0  K^-)\,,
\end{equation}
where
\begin{equation}
\bar \epsilon_2 (s) = \frac{1}{16 \pi}\
\frac{1}{\cos (\pi\alpha_0/2)}\
\frac{e^{-i \pi\alpha_0/2}}{s_0\alpha'
\left( \ln (s/s_0) - i \pi /2 \right)}\,.
\end{equation}
We restore the FSI contribution to
$A (B^-\to \overline{K^0} \pi^-)$ by use of a dispersion relation.  The dispersion
integral may be evaluated in closed form with the approximations
$\alpha_0={1\over2}$ and $\ln(s/s_0)=\ln(m_B^2/s_0)$,
\begin{eqnarray}
A (B^- \to \overline{K^0} \pi^-)_{\rm FSI} &=&
\gamma\,\bar \epsilon_2 (m_B^2)\,A (B^- \to
\pi^0 (\eta) K^-)\times{1\over\pi} \int_{(m_\pi+m_K)^2}^\infty
\frac{d s}{s-m_B^2} \left ( \frac{s}{s_0} \right )^{\alpha_0 - 1}
\nonumber \\
&=& i\,\gamma\,\bar \epsilon_2 (m_B^2)\, \frac{\sqrt{s_0}}{m_B}
A (B^- \to \pi^0 K^-)\,,
\end{eqnarray}
where we have taken the limit $m_{\pi(K)}^2/m_B^2\to0$.
For rescattering through the $\eta K^-$ channel, the only difference is an
$SU(3)$ group theory factor in the residue function $\gamma$.  The magnitude
of the contribution of a given channel to the soft rescattering amplitudes
defined in the previous section is then
\begin{equation}\label{eps2}
\epsilon_2=\gamma\,{\sqrt{s_0}\over m_B}|\bar\epsilon_2(m_B^2)|\,,
\end{equation}
where the residue $\gamma$ depends on the channel.

Finally, we must make a numerical estimate of $\gamma$, which parameterizes
the coupling of the $\rho$ trajectory to the pseudoscalar mesons.  These
couplings may be estimated by considering $pp$ and $\pi^+ p$ scattering data
and using $SU(3)$ relations.  The exchange of the $\rho$ trajectory requires a
coupling at each vertex, so
\begin{equation} \label{res}
\gamma_\rho (pp \to pp) = \gamma_{p p \rho}^2,\qquad\qquad
\gamma_\rho (\pi p \to \pi p) = \gamma_{\pi\pi\rho}
\gamma_{p p \rho}\,.
\end{equation}
The optical theorem gives
\begin{equation} \label{optical}
\sigma_{tot} = \frac{1}{s} {\cal M}_{f \to f}
\end{equation}
for both $pp$ and $\pi p$ scattering. The forward scattering
amplitude is obtained from Eq.~(\ref{reggeme}), written for $pp$ and
$\pi p$ scattering, by setting $t = 0$ (that is, $\cos \theta = 1$). The
residue functions (\ref{res}) entering the expression for the forward
scattering amplitudes are fixed from the Particle Data Group parameterizations
of $pp$ and $\pi p$ scattering data~\cite{PDG},
\begin{equation} \label{elast}
\sigma_{\rm tot}^{ik} = X_{ik} \left ( \frac{s}{s_0} \right )^{0.08} +
Y_{ik} \left ( \frac{s}{s_0} \right )^{-0.56}\,.
\end{equation}
The first term represents the Pomeron contribution.  The second comes from
the $\rho$ trajectory for $\pi^+p$ scattering, and is a combined contribution
from the $\rho$ and $a$ trajectories for $pp$ scattering.  The Particle Data
Group
fit gives $Y_{pp} = 56.08\,$mb and $Y_{\pi^+ p} = 27.56\,$mb.  Assuming that
the $\rho$ and $a$ trajectories contribute equally in the $pp$ channel, and
using again the approximation $\alpha_0={1\over2}$,
we find
\begin{equation}
\gamma_0^2\equiv\gamma_{\pi\pi\rho}^2={2s_0 Y^2_{\pi p}\over Y_{pp}}
\approx72\,.
\end{equation}
As defined, $\gamma_0$ is the $\pi^+\pi^-\rho^0$ coupling, since of the
vector meson octet, only the $\rho_0$ contributes in the $\pi^+ p$ channel.
The residue functions which we will require may be found by applying $SU(3)$
symmetry.  For $\rho$ exchange in the $\pi K$ channel, we find
$\gamma_{\pi\pi\rho}=\gamma_0$, $\gamma_{KK\rho}=-\gamma_0/\sqrt2$, and
$\gamma_{\pi\eta\rho}=0$.  As a consequence, we have for the $\rho$
trajectory,
\begin{eqnarray}\label{rhoresidues}
\gamma_\rho(\pi^0 K^-\to\pi^- \overline{K^0})&=&-\frac{1}{\sqrt2}\,\gamma_0^2\,,
\nonumber\\
\gamma_\rho (\eta K^- \to \pi^- \overline{K^0}) &=& 0\,,
\end{eqnarray}
and similarly for $K^*$ exchange,
\begin{eqnarray}\label{Kresidues}
\gamma_{K^*}(\pi^0 K^-\to\pi^-
\overline{K^0})&=&-{1\over2}\sqrt{1\over2}\,\gamma_0^2\,,\nonumber\\
\gamma_{K^*}(\eta K^-\to\pi^-
\overline{K^0})&=&{1\over2}\sqrt{3\over2}\,\gamma_0^2\,.
\end{eqnarray}  With these residues, we complete the computation of the
rescattering amplitude $\epsilon_2$.

We are now in a position to estimate the contribution to $\epsilon$ from a
given two body rescattering channel, by combining the rescattering amplitudes
(\ref{eps2}), the production amplitude ratios (\ref{ratios}), and the residues
(\ref{rhoresidues}) and (\ref{Kresidues}).  For example, for the intermediate
$\pi^0K^-$ state, rescattering via $\rho$ exchange, we find a term
\begin{equation}
  \epsilon_{\pi K\rho}=\left({A_{us}^{n+} \over A_{cs}^+
  }\right)_{\pi^0}{\sqrt{s_0}\over m_B}
  \, |\bar\epsilon_2(m_B^2)|\,\gamma_\rho(\pi^0
K^-\to\pi^- \overline{K^0})\,.
\end{equation}
We do not know the strong phase which multiplies $\epsilon_{\pi
K\rho}$.  In adding the three contributions $\epsilon_{\pi K\rho}$,
$\epsilon_{\pi KK^*}$, and $\epsilon_{\eta KK^*}$ (recall that $\epsilon_{\eta
K\rho}$ vanishes), we must make some assumption about their relative 
phases.\footnote{We prefer not to impose constraints from $SU(3)$ symmetry on the 
phases because of the substantial model-dependence already present in our calculation.
Doing so would not change substantially the magnitude of the rescattering effects, but
it would imply a smaller uncertainty then we would advocate.} 
For the purpose of this estimate we might imagine adding them incoherently, so
$\epsilon^2 = \epsilon^2_{\pi K\rho}+ \epsilon^2_{\eta KK^*}+ \epsilon^2_{\eta
K\rho}$.  Our estimates for the various channels are
$\epsilon_{\pi K\rho}\simeq0.044$, $\epsilon_{\pi KK^*}\simeq0.022$, and
$\epsilon_{\eta KK^*}\simeq0.022$, which by this prescription would give
$\epsilon\sim0.06$.  Alternatively, adding them coherently would yield
$\epsilon\sim0.09$.  These unknown relative strong phases are one important source
of uncertainty in this estimate of $\epsilon$.

Of course, there are many other sources of uncertainty as well.  Perhaps the most
severe of these is the neglect of multibody intermediate   states, which we have
omitted because we have no good model for them.  For this reason, we are likely
to have, if anything, underestimated the effect of FSI.  More model-dependence
arises from the use of factorization to estimate the ratios $A_{us}^{2+}
/A_{cs}^+$, given that this ansatz is known at times to
fail in $B$ decays to light pseudoscalar mesons.  Smaller uncertainties
arise from the phenomenological extraction of the residues and from the use of
$SU(3)$ symmetry. Finally, we note that in some (but not all) models of Regge
exchange, there may be cancelations which further suppress the small contribution
from the $K^*$ exchange to both the $\eta K^-$ and $\pi^0 K^-$
intermediate states. (For recent discussions of these questions, see
Refs.~\cite{Lipkin,DoLa,FFP}.) Similar effects may enhance the contribution
from $\eta^\prime K^-$.\footnote{We are grateful to H.~Lipkin for
discussions of this point, and for stressing to us the important role
played by tensor meson exchange.}  Our neglect of the
$\eta^\prime$ has been conservative, in the sense that it has likely
caused us to underestimate the total effect of rescattering.  We have
done so because, in the presence of the anomaly, the $\eta^\prime$ is not
related by unitary symmetry to any other meson.  Hence we have little
guidance, other than from the quark model, for how to include it.
In the end, it certainly would be unwise to trust our
estimate of $\epsilon$ to better than a factor of two, and even that
much confidence would be optimistic.  The same caveat should be applied to {\it
any\/} phenomenological model of soft final state interactions.  What is important
here is that we have found neither a dominant effect of order one or larger, nor an
insignificant effect of order one percent.

%%%%%%%%%%%%%%%%%%%%%%%%%%%%%%%%%%%%%%%%%%%%%
\subsection{The effect of FSI on $\Adir$ and $R$}

We close by returning to the effect of FSI on the observables
$\Adir$ and
$R$.  We have seen from our model that rescattering effects as large as
$\epsilon\sim0.1-0.2$ easily could be consistent with the Standard Model.
Therefore, unless $\gamma$ were known independently to be very small, an observation
of $\Adir\sim0.2$ could {\it not\/} be taken unambiguously to be
a signal for New Physics.  Note that we do not claim to predict such a large
asymmetry; we simply observe that it would be neither unnatural nor surprising for
it to be generated by final state interactions.  On the other hand, we would
maintain that a larger asymmetry, such as $\Adir\sim{\cal O}(1)$,
would still be an exciting sign of a source of CP violation beyond the CKM matrix.

Similarly, we can consider the effect of $\epsilon\sim0.1$ on the bounds
(\ref{corrR}) and (\ref{FMmod}) on $\gamma$.  For example, the fractional
correction to the bound on
$|\cos\gamma|$ is $\Delta\equiv \epsilon R/\sqrt{1-R}$.  The value of $\Delta$ is a
strong function of the experimentally observed $R_{\rm exp}$,
\begin{equation}
\begin{array}{l}
\Delta\simeq \epsilon ~~\mbox{for}~~ R_{\rm exp} = 0.65\,, \\
\Delta\simeq 2\epsilon ~~\mbox{for}~~ R_{\rm exp} = 0.80\,.
\end{array}
\end{equation}
The bound deteriorates quickly as $R_{\rm exp} \to 1$.  In terms of $\sin^2\gamma$,
we certainly would conclude that the observation $\sin^2\gamma\sim1.2R_{\rm exp}$
would {\it not\/} constitute an unambiguous signal of New Physics.

%%%%%%%%%%%%%%%%%%%%%%%%%%%%%%%%%%%%%%%%%%%%%
\section{Model Independent Bounds on the FSI Corrections}

Our phenomenological model suggests that
soft FSI contributions of the current-current operators
$Q_{1,2}^{us} $ could account
for ${\cal O}(10\%)$ of the $B^+ \to K^0 \pi^+ $ amplitude.
This has a dramatic consequence, namely making
$\Adir \sim 10-20\% $ realistic within the
Standard Model. In view of the large theoretical uncertainties involved, it would be
extremely useful to find an experimental method by which to bound the magnitude of
the FSI contribution. The observation of a larger asymmetry would then be a signal
for New Physics. In this Section we describe such an attempt, along the lines
proposed  in Ref.~\cite{giw} for the decay $B \to \phi K $.
The idea is to find decay modes mediated by the quark level transition $b \to s \bar
s d$, for which branching ratio measurements or upper bounds would, by
application of  flavor $SU(3)$ flavor symmetry, imply a direct upper bound on
$\epsilon$.

The most interesting modes in our case turn out to be $B^\pm \to K^\pm K $.
The effective Hamiltonian for $b \to d$ transitions may be obtained from
(\ref{HeffCP}) by the substitution of $s \to d$ in the operators and in the
indices of the CKM matrix elements.  In analogy with Eq.~(\ref{Camps}) the
amplitudes
may be decomposed according to  their dependence on CKM factors, giving
\beqa \label{KKamps}
A(B^+ \to K^+ \bKz ) &=& A_{cd} - A_{ud} e^{i \gamma} e^{i \delta}\,,\nonumber\\
A(B^- \to K^- K^0 ) &=& A_{cd} - A_{ud} e^{- i \gamma} e^{i \delta}\,. \eeqa
Invariance under the $SU(3)$ rotation
$\exp(i {\pi \over 2} \lambda_7 )$, i.e., interchange of $s$ and $d$ quark fields,
implies equalities among operator matrix elements,
\beqa \label{su3} \langle K^- K^0 | Q_i^{qd} | B^- \rangle & = &
\langle K^0 \pi^- | Q_i^{qs} | B^- \rangle,~~~q=u,c,;~~i = 1,2 \\
\langle K^- K^0 | Q_i^{d} | B^- \rangle & = &
\langle K^0 \pi^- | Q_i^{s} | B^- \rangle, ~~~i = 3,..,10\,. \nonumber \eeqa
These lead to the relations
\beq\label{su3A}  A_{ud} e^{i \delta} = A_{us}^+ e^{i \delta_+}
{V_{ud} \over V_{us} } (1 + R_{ud} )\,,
~~~A_{cd} = A_{cs}^+ {V_{cd} \over V_{cs} } (1 + R_{cd} )\,,\eeq
where $R_{ud}$ and $R_{cd}$ parameterize $SU(3)$ violation, which
is typically of the order of $20-30\%$.  Note that it is only an $SU(2)$ subgroup
of $SU(3)$, namely $U$-spin, which is required to derive these relations.  Since
the $B^-$ carries $U=0$ and the transition operators $Q_i^{qd}$ and $Q_i^d$
carry $U={1\over2}$, it is only the $U={1\over2}$ component of the $K^-K^0$
final state which couples to the decay channel.  As a result, we have the freedom to
add any pure $U={3\over2}$ combination to the final state, involving additional
$\pi^-\pi^0$ and $\pi^-\eta$ pairs, without affecting the relations (\ref{su3}).
As it turns out, it is the simple combination $K^-K^0$ which yields the most
phenomenologically interesting bound.

An upper bound on $\epsilon$ follows from the ratio
\beq\label{RK} R_K = {BR(B^+ \to K^+ \overline{K^0} ) +
BR(B^- \to K^- K^0 ) \over BR(B^+ \to K^0
\pi^+ ) +BR(B^- \to \overline{K^0}  \pi^- )}\,. \eeq
After some algebra, we obtain
\beq\label{rmax}\label{epsbound}
\epsilon < \lambda \sqrt{R_{K}}\, (1 + {\rm Re}[R_{ud} ]) + \lambda^2 (R_{K} +
1 ) \cos\gamma \cos \delta_+  + {\cal O}(\lambda^3 ,
\lambda^2 R_{ud,cd} )\,,
\label{eq:rmax} \eeq
where $\lambda\simeq0.22$ is the Wolfenstein parameter.  (In deriving this relation,
we assume that the full matrix element for $B^-\to K^-K^0$ is not smaller, in
magnitude, than the partial contribution from rescattering.  It is
possible, if there is some fine tuned cancelation, for
rescattering actually to lower the branching ratio; in this case, the
relation~(\ref{epsbound}) would not be valid.)  This bound becomes more reliable if
we set
$\cos\gamma\cos\delta_+ = 1$.  With Eq.~(\ref{FMmod}), we find a bound on $\gamma$,
\beq\label{gammamod}\sin^2 \gamma < R +  2\lambda R \sqrt{R_K (1 - R)}
+ {\cal O}(\lambda^2 ,\lambda R_{ud,cd} )\,, \eeq
ignoring electroweak penguin operators.  One could include them, as before,
with the substitution in~(\ref{EWsub}).

We are also interested in obtaining an upper bound on $\Adir$.
Keeping $\cos\delta$ free gives
\beq |\Adir| < 2 \epsilon \sin \gamma + {\cal O}(\epsilon^3 ) .
\eeq In the absence of a nontrivial bound on $\gamma$ from Eq.~(\ref{gammamod}),
$\Adir$ is maximized at $\sin \gamma = 1 $,
leading to
\beq |\Adir| < 2 \lambda \sqrt{R_{K}}\, (1 + {\rm Re}[R_{ud}] )
+ {\cal O}(\lambda^3 , \lambda^2 R_{ud,cd} )\,. \eeq
However, an independent bound on $\gamma$
would place a tighter limit on the asymmetry,
\beq |\Adir| < 2 \lambda \sqrt{R_{K} R}\, (1 + Re[R_{ud}])+ 2
\lambda^2 \sqrt{R}\, (R_K\sqrt{1-R}+R_{K} + 1)
+ {\cal O}(\lambda^3 , \lambda^2 R_{ud,cd} )\,.\eeq
Again, electroweak penguin operators may be included with the substitution
(\ref{EWsub}).  Following Ref.~\cite{giw}, analogous bounds on
$\Adir (B^\pm \to
\phi K^{(*)\pm}) $ can be obtained by substituting for $R_K$ the ratio
$\left [ \sqrt{BR(B^\pm \to K^{0(*)} K^\pm )} + 
\sqrt{BR(B^\pm \to \phi \pi^\pm(\rho^\pm))} \right ] / \sqrt{BR(B^\pm \to \phi K^\pm )}$.

The CLEO Collaboration has recently obtained the upper bound
$R_K < 0.95 $ at 90\% C.L., including only statistical errors,
and approximately $R_K < 1.9 $ once systematic errors are
included as well~\cite{privcomm}.   Unfortunately, this is not very restrictive,
implying $\epsilon < 0.4 $ and $\Adir < 0.6 $.
It is possible that more interesting constraints on $\epsilon$ and $\Adir$ could be
obtained in the future, given that $R_K \sim {\cal O}(\lambda^2)$ in the limit of
vanishing FSI. Ultimately, the utility of such a bound is limited by the fact
that, for $\epsilon$ small enough, $R_K$ becomes independent of $\epsilon$, since
rescattering channels are then negligible compared with other contributions.
Conversely, a large observed value for $R_K$ would not necessarily mean that FSI
contributions are large, since this could also result from New Physics which
enhances
the $b \to d$ transitions. Resolving the question of how to distinguish these two
possibilities is left for the future.

%%%%%%%%%%%%%%%%%%%%%%%%%%%%%%%%%%%%%%%%%%%%%
\section{New Physics}

Assuming that a large CP asymmetry is observed in $B^\pm\to\pi^\pm K$
decays or that $\sin\gamma$ is measured independently and found to violate
the bound~(\ref{bound}), this could be explained either by large
soft rescattering effects or by New Physics. From our calculations above,
it is clear that theory cannot at present exclude
the possibility of large FSI effects. Yet, it is also possible that these
effects are not large and that, in the future, the experimental tests
proposed in the previous section will imply that New Physics indeed is
required. It is important, then, to understand which extensions of the Standard Model
could contribute significantly (and with new CP violating phases) to the relevant
$B\to\pi K$ modes. Furthermore, we would like to understand whether
New Physics contributions to the charged and neutral modes should
be expected to have any special features (such as isospin symmetry
relations) which will allow further tests.

The most general New Physics effects on the amplitudes (\ref{Camps})
can be parameterized
by six new parameters,
\beqa\label{fourmodes}
A(B^0\to\pi^- K^+)&=A^0_{\rm SM}-A_N^u e^{i\phi_u}e^{i\delta_u},\ \ \
A(\bBz\to\pi^+K^-)&=\bar A^0_{\rm SM}-A_N^ue^{-i\phi_u}e^{i\delta_u},
\nonumber\\
A(B^+\to\pi^+ K^0)&=A^+_{\rm SM}-A_N^d e^{i\phi_d}e^{i\delta_d},\ \ \
A(B^-\to\pi^-\bKz)&=A^-_{\rm SM}-A_N^d e^{-i\phi_d}e^{i\delta_d}.
\eeqa
Here $A_{\rm SM}$ are the Standard Model amplitudes, where $A^0$ and
$\bar A^0$ carry the same strong phases and opposite weak phases,
as do $A^\pm$.  We introduce the New Physics amplitudes $A_N^{u,d}$, with CP
violating phases $\phi_{u,d}$ and strong phases $\delta_{u,d}$.

A first class of models are those which give potentially large
tree-level contributions to $\Delta B = 1$ four quark operators.
These include supersymmetry without $R$ parity and models with extraquarks in vector-like representations of the SM gauge group. A second class
of models gives new contributions to four quark operators through loopdiagrams only, with new particles running in the loop.
This includes various supersymmetric flavor models and a sequential
fourth generation. A third class of models gives new contributions tothe $\Delta B =1 $ chromomagnetic dipole operators.
Finally, there are models where no significant effect is expected.
These include models of extra scalars and Left-Right Symmetric models.

\subsection{Supersymmetry without R Parity}

In supersymmetric models without $R_p$ there are new, slepton
mediated tree diagrams, contributing at tree level to the
$b\to u\bar us$ and $b\to d\bar ds$ transitions.

First, consider the $\lambda^\prime_{ijk}L_iQ_j\bar d_k$ couplings.
For $b\to u\bar us$ transitions, the contributions come from charged
slepton mediated diagrams, while for $b\to d\bar ds$ transitions,
the contributions come from sneutrino mediated diagrams,
\beqa\label{lpcon}
A_N^u e^{i\phi_u}&\propto&\sum_{i=1}^3{\lambda^\prime_{i13}
\lambda^{\prime*}_{i12}\over m^2(\tilde\ell^-_i)}\,,\\
A_N^d e^{i\phi_d}&\propto&\sum_{i=1}^3{
\lambda^\prime_{i13}\lambda^{\prime*}_{i12}+
\lambda^\prime_{i23}\lambda^{\prime*}_{i11}+
\lambda^\prime_{i21}\lambda^{\prime*}_{i31}+
\lambda^\prime_{i11}\lambda^{\prime*}_{i32}
\over m^2(\tilde\nu_i)}\,.
\eeqa
We learn that indeed the new physics introduces six new independent
parameters. In the special case where $(i)$ $m^2(\tilde\ell_i^-)=
m^2(\tilde\nu_i)$ and $(ii)$ $\lambda^\prime_{i13}\lambda^{\prime*}_{i12}$
is much larger than the other three combinations that appear in (\ref{lpcon}),
isospin is a good symmetry (similar to the SM QCD penguin diagrams).
The first condition is fulfilled in many models, but the second
is not. Generically, we have $A_N^d\neq A_N^u$, $\phi_d\neq\phi_u$
and $\delta_u\neq\delta_d$.

Second, consider the $\lambda^{\prime \prime}_{ijk}\bar u_i\bar d_j\bar d_k$ couplings
and note that these couplings are antisymmetric in $(j,k)$.
For $b\to u\bar us$ transitions, the contributions come from the down
squark mediated diagrams only, while for $b\to d\bar ds$ transitions,
the contributions come from up, charm and top squark mediated diagrams:
\beqa\label{lppcon}
A_N^u e^{i\phi_u}&\propto&{\lambda^{\prime \prime}_{113}
\lambda^{\prime \prime}_{112}
\over m^2(\tilde d_1)}\,,\\
A_N^d e^{i\phi_d}&\propto&\sum_{i=1}^3{
\lambda^{\prime \prime}_{i13} \lambda^{\prime \prime *}_{i12}
\over m^2(\tilde u_i)}\,.
\eeqa
Again, there are six new independent parameters unless
$(i)$ the $i=1$ contribution dominates $A_N^d$ and
$(ii)$ $m^2(\tilde d_1)=m^2(\tilde u_1)$, in which case isospin is a good
symmetry. The first condition is unlikely to be fulfilled.
We expect $A_N^d\neq A_N^u$, $\phi_d\neq\phi_u$
and $\delta_u\neq\delta_d$.

%%%%%%%%%%%%%%%%%%%%%%%%%%%%
\subsection{Singlet Down Quarks}

Models with additional $SU(2)_L$-singlet down quarks could lead to
dramatic effects in CP violation in the interference between decays with
and without mixing~\cite{NiSi}.
This is a result of $Z$-mediated tree level contributions  to
$B-\overline B$ mixing. Could there also be large effects in $\Adir$?

Obviously, there are new contributions to the decay amplitudes
from $Z$-mediated tree level diagrams,
\beqa\label{Zmed}
A_N^u e^{i\phi_u}&\propto&{U_{sb}g_{Zuu}\over m_Z^2}\,,\\
A_N^d e^{i\phi_d}&\propto&{U_{sb}g_{Zdd}\over m_Z^2}\,,
\eeqa
where $g_{Zff}\propto (T_3^f-Q_f\sin^2\theta_W)$, and $U_{sb}$
is the mixing angle in the $Z$ couplings.
The present experimental bound on $BR(B\to X\mu^+\mu^-)$ gives
$|U_{sb}/(V_{cb}V_{cs})|\leq0.04$. This suppression roughly
compensates for the loop suppression of the standard model
QCD penguin. Thus, we learn that

$(i)$ $A_N$ could be of the same order as $A_{\rm SM}$;

$(ii)$ $A_N^u\neq A_N^d$  because of the $g_{Zqq}$ factor;

$(iii)$ $\phi_u=\phi_d=\arg(U_{sb})$.

\subsection{Supersymmetric Flavor Models}

Supersymmetric models (with $R_p$ conserved) contribute to
$b\to s$ transitions through penguin diagrams with squark - gluino
loops. The ratio between this contribution and the standard model
penguin amplitude may be estimated to be \cite{BaSt}
\beq\label{GenSUSY}
{A^{\rm SUSY}\over A^{\rm SM}}\sim{\alpha_3\over\alpha_2}
{m_W^2\over{\rm max}(m^2_{\tilde b},M_3^2)}
{K_{32}^*K_{33}\over V_{tb}V_{ts}^*}{\eta\over\ln(m_t^2/m_c^2)}\,,
\eeq
where $K$ is the mixing matrix for the gluino couplings and
$\eta\sim(m^2_{\tilde b}-m^2_{\tilde s})/(m^2_{\tilde b}+m^2_{\tilde s})$
is a measure of the non-universality in the squark masses.
We see that even in the absence of a super-GIM mechanism, namely
with $\eta\sim1$, the supersymmetric contribution is comparable
to the Standard Model one only if the relevant mixing angle, namely
$K_{32}^*K_{33}$, is large. Such a situation is phenomenologically
allowed. Furthermore, the fact that $m_s/m_b\sim|V_{cb}|$ implies that a large
$\tilde b_R - s_R$ mixing ($K_{32}\sim{\cal O}(1)$)
is not unlikely \cite{LNSa}. Below we examine whether this
possibility is indeed realized in various supersymmetric
flavor models.

Most flavor problems of supersymmetry are solved (without giving up
naturalness) in models where the first two squark generations are heavy
\cite{DKS,DKL,PoTo,CKN,CKLN}. Mixing angles with the third generation
can be large. In these models, for $\tilde b_L$ and
gluino masses in the range of 100 -- 300~GeV, it was found \cite{GrWo}
that the supersymmetric QCD penguins (namely, squark - gluino loops)
can be twice as large as the standard model ones for $b\to s$
transitions, and can carry a new phase. Similar to
the standard model QCD penguins, isospin is a good symmetry here,
namely $A_N^d=A_N^u$, $\phi_d=\phi_u$ and $\delta_d=\delta_u$.

The situation is different in models of Abelian
horizontal symmetries, where alignment of quark and squark mass
matrices is the mechanism which suppresses supersymmetric
contributions to flavor-changing neutral currents (FCNC)~\cite{NiSe,LNSb}. The
constraints from
$K-\overline K$ mixing require that the $(32)$ entries in the
mass matrices are also small, leading to a suppression of $K_{32}$.
A similar situation occurs
in models of non-Abelian horizontal symmetries, where degeneracy
of the first two squark generations suppresses FCNC \cite{DKL,BHR,CHM,HaMu}.
Typically, the
first two generations are in a doublet and the third in a singlet
of the horizontal symmetry. Then, $\tilde b_{L,R}-s_{L,R}$
mixing does break the horizontal symmetry and is, therefore, suppressed:
$K_{32}K_{33}$ is of order $V_{ts}V_{tb}$ and the supersymmetric
contribution to the $b\to s$ transition is typically small
\cite{BaSt,CFMMS}.

Finally, if CP is an approximate symmetry of the New Physics (which is a viable
possibility in the supersymmetric framework), so that all CP violating
phases are small, say $\phi_{u,d}={\cal O}(10^{-3})$, then we expect New Physics to
contribute to $\Adir$ at the level $\le 10^{-3}$ regardless of the size of the
supersymmetric contributions to the various $B\to\pi K$ decays.

%%%%%%%%%%%%%%%%%%%%%%%%%%%%%%%%%%%%
\subsection{Fourth Quark Generation}

Models of four quark generation require, of course, that the
fourth generation neutrino is rather heavy ($\geq m_Z/2$).
If this possibility is realized in nature, then we expect
large new contributions from QCD penguin diagrams with
$W$ - $t^\prime$ loops. The $m_{t^\prime}$-dependence of this
contribution may compensate for a possibly small CKM factor,
$V_{t^\prime b}V_{t^\prime s}^*$, and become a significant, if not
dominant, contribution. Furthermore, as the $4\times 4$ quark mixing
matrix has three independent CP violating phases, this contribution is
likely to carry a new phase. Isospin should be a good symmetry
and $\delta_u=\delta_d$ is likely.

%%%%%%%%%%%%%%%%%%%%%%%%%%%%%%%%%%%%
\subsection{Models with Enhanced $b \to sg$.}

Models with enhanced $b \to sg$ dipole operator coefficients, leading to
${\cal B}(b \to sg) \sim 10\% $, have been suggested as a
possible resolution of several potential puzzles in inclusive
$B$ decays \cite{hou,kagan}. Examples have been discussed which employ
squark-gluino loops, vectorlike quark-neutral scalar loops, or
techniscalar exchange \cite{kagan,ciuchinidipole}.
Despite the large overall rate, the dipole induced amplitudes for
rare $B$ decays are of same order as the Standard Model amplitudes and so
the two can interfere substantially. Arbitrary new weak phases in the dipole
operator coefficients can therefore lead to sizable $A_{CP}^{dir} \ge 10\% $
\cite{hawaii}. More modest enhancement of $b \to sg$ can also lead to sizable $CP$
asymmetries. Such an example has been discussed in the context of topcolor models
\cite{valencia}. Again isospin is a good symmetry.

%%%%%%%%%%%%%%%%%%%%%%%%%%%%%%%%%%%
\subsection{Discussion}

There exist, of course, extensions of the standard model where
large new contributions to the relevant $B\to\pi K$ decays are
unlikely. Below we give a few examples.

Charged Higgs mediated tree diagrams
contribute only to $A_N^u$, so $A_N^d=0$.
The CKM combinations are similar to those in the standard model
$W$ mediated tree diagrams, namely we still have
$\phi_u=\gamma$. The contribution is, however, small
compared to the Standard Model $A_{us}^+$ because, while the CKM suppression
persists, we now have in addition a strong suppression from small Yukawa
couplings. Consequently, there is no effect on $\Adir$ or on $R$.

Neutral Higgs exchange in models without natural flavor conservation
(NFC) could contribute to both $b\to u\bar us$ and $b\to d\bar d s$.
But if the smallness of scalar mediated FCNC is explained by
a horizontal symmetry, then we expect a suppression of order
$\leq{\cal O}(m_bm_{u,d}/m_Z^2)\sim10^{-5}$, which  means that
the effects are negligible.

In left-right symmetric models there is a new contribution from
$W_R$-mediated decays, but it is suppressed by
${\cal O}(m_{W_L}^2/m_{W_R}^2)\leq10^{-2}$.
The CKM ratio is expected to be ${\cal O}(1)$.
So, again, we expect no observable effects on either $\Adir$ or $R$.

To summarize the situation regarding new physics effects,
we note the following points:

$(i)$ There are several well-motivated extensions of the standard model
which can significantly affect $B\to\pi K$ decays.

$(ii)$ In models where there are new tree diagram contributions,
there are no relations, in general, between the contributions
to $B^+\to\pi^+ K^0$ and the contributions to $B^0\to\pi^- K^+$.
The new physics effects in the amplitudes (\ref{Camps}) introduce six new
parameters.

$(iii)$ In models where the new contributions are through QCD penguin or 
chromomagnetic dipole operators, isospin is a good symmetry, and the number of new 
parameters is therefore reduced from six to three.

$(iv)$ We note that if a large CP asymmetry (\ref{ABpiK}) is measured,
that will invalidate the bound~(\ref{bound}). In contrast, if a
small CP asymmetry is measured, that would not provide an
unambiguous confirmation for the validity of this
bound, because it could be a result of small strong phases
rather than a small magnitude of final state interactions.

%%%%%%%%%%%%%%%%%%%%%%%%%%%%%%%%%%%%%%%%%%%%%%%%%%%
\section{Conclusions}

We have analyzed the effect of final state interactions on the search for New
Physics in $B\to\pi K$ decays.  Using a phenomenological model, we found that while
such effects are unlikely to be large enough to dominate individual branching
fractions, they can still complicate those avenues for identifying New Physics
which rely on a Standard Model suppression of weak phases in the matrix elements
for $B^\pm\to\pi^\pm K$.  As a result, and in contrast to previous expectations,
we conclude that the observation of $\Adir\sim0.2$ or $\sin^2\gamma\sim1.2R$ would {\it
not\/} be an unambiguous sign of a source of CP violation beyond the CKM matrix.
While we do not claim to compute the magnitude of final state interactions reliably,
our model is sufficient to demonstrate that effects of this size are entirely
generic and cannot be ruled out without independent {\it empirical\/} evidence.  We
propose a simple test which could probe this question experimentally.  Finally, in
anticipation of the observation of CP violation in these channels at a level which
cannot be explained by the Standard Model, we discuss the features of various
models of New Physics as they would be manifested in these decays.

%%%%%%%%%%%%%%%%%%%%%%%%%%%%%%%%%%%%%%%%%%%%%%%%%%%
\acknowledgments

We thank R.~Fleischer, H.~Lipkin, T.~Mannel, M.~Neubert, M.~Peskin, H.~Quinn,
A.~Schwimmer and M.~Worah for useful conversations.  A.F.~and A.P.~were supported in part by the United States National Science  Foundation under Grant
No.~PHY-9404057. A.F.~was also supported by the United States National Science
Foundation under National Young Investigator Award No.~PHY-9457916, by the United
States Department of Energy under Outstanding Junior Investigator AwardNo.~DE-FG02-94ER40869, and by the Alfred P.~Sloan Foundation.
A.F. is a Cottrell Scholar of the Research Corporation. Y.N. is supported in part
by the United States--Israel Binational Science Foundation (BSF), by the Israel
Science Foundation, and by Minerva Foundation (Munich). A.K. was supported by the
United States Department of Energy under Grant No.~DE-FG02-84ER40153.
A.K. would like to thank the Weizmann Institute and the CERN Theory group for
their hospitality while this work was in progress.

%%%%%%%%%%%%%%%%%%
\tighten

\end{document}